%% file: main.tex
\documentclass[conference]{IEEEtran}
\usepackage{amsmath}
\usepackage{amssymb}
\usepackage{graphicx}
\usepackage{url}
\usepackage{tikz}
\usepackage{pgfplots}
\usepackage{float}
\pgfplotsset{compat=1.17}
\pgfplotsset{every axis/.append style={
    label style={font=\footnotesize},
    tick label style={font=\footnotesize},
    legend style={font=\scriptsize}
}}
\usepackage[hidelinks]{hyperref}
\usepackage{orcidlink}
\usepackage{balance}
\usepackage{xcolor}
\usepackage{booktabs}
\usepackage[acronym,nomain]{glossaries}

\newacronym{afdm}{AFDM}{affine frequency division multiplexing}
\newacronym{daft}{DAFT}{discrete affine Fourier transform}
\newacronym{daftsafdm}{DAFT-s-AFDM}{discrete affine Fourier transform-spread affine frequency division multiplexing}
\newacronym{ofdm}{OFDM}{orthogonal frequency division multiplexing}
\newacronym{dft}{DFT}{discrete Fourier transform}
\newacronym{dftsofdm}{DFT-s-OFDM}{discrete Fourier transform-spread orthogonal frequency division multiplexing}
\newacronym{nr}{NR}{New Radio}
\newacronym{afdma}{AFDMA}{affine frequency division multiple access}
\newacronym{papr}{PAPR}{peak-to-average power ratio}
\newacronym{ber}{BER}{bit error rate}
\newacronym{isac}{ISAC}{integrated sensing and communications}
\newacronym{mmse}{MMSE}{minimum mean-square error}
\newacronym{idft}{IDFT}{inverse discrete Fourier transform}
\newacronym{idaft}{IDAFT}{inverse discrete affine Fourier transform}
\newacronym{qpsk}{QPSK}{quadrature phase-shift keying}
\newacronym{cpp}{CPP}{chirp-periodic prefix}
\newacronym{awgn}{AWGN}{additive white Gaussian noise}
\newacronym{dfnt}{DFnT}{discrete Fresnel transformation}
\newacronym{mse}{MSE}{mean-square error}
\newacronym{sinr}{SINR}{signal-to-interference-plus-noise ratio}
\newacronym{ccdf}{CCDF}{complementary cumulative distribution function}
\newacronym{imt}{IMT}{International Mobile Telecommunications}
\newacronym{fiveg}{5G}{fifth-generation}
\newacronym{sixg}{6G}{sixth-generation}
\newacronym{snr}{SNR}{signal-to-noise ratio}
\glsdisablehyper

\ifdefined\cleanversion
  
\else
  
\fi

\newtheorem{proposition}{Proposition}
\begin{document}

\title{Effects of Residual Chirp Rate on \glsentryshort{daftsafdm} Uplink Performance}

\author{%
\IEEEauthorblockN{%
Henrique L. Senger\textsuperscript{\orcidlink{0009-0004-1586-8168}}\IEEEauthorrefmark{1},
Gustavo P. Gonçalves\textsuperscript{\orcidlink{0009-0000-8260-4390}}\IEEEauthorrefmark{1},
Bruno S. Chang\textsuperscript{\orcidlink{0000-0003-0232-7640}}\IEEEauthorrefmark{1},\\
Hyeon Seok Rou\textsuperscript{\orcidlink{0000-0003-3483-7629}}\IEEEauthorrefmark{2}, and
Giuseppe Thadeu Freitas de Abreu\textsuperscript{\orcidlink{0000-0002-5018-8174}}\IEEEauthorrefmark{2}%
}
\IEEEauthorblockA{%
\IEEEauthorrefmark{1}CPGEI/Electronics Department,
Federal University of Technology -- Paraná, Curitiba, Brazil\\[0.15ex]
\IEEEauthorrefmark{2}School of Computer Science and Engineering,
Constructor University, Bremen, Germany\\[0.15ex]
Emails: \{hsenger, gustavog.1999\}@alunos.utfpr.edu.br,  bschang@utfpr.edu.br, \\ \{hrou, gabreu\}@constructor.university%
}
}

\maketitle

\begin{abstract}
Affine frequency division multiplexing (\glsentryshort{afdm}) is robust to doubly dispersive channels in high-mobility \glsentryshort{sixg} systems but, like orthogonal frequency division multiplexing (\glsentryshort{ofdm}), suffers from a high peak-to-average power ratio (\glsentryshort{papr}). Spreading with a discrete affine Fourier transform (\glsentryshort{daft}) yields \glsentryshort{daftsafdm}, whose precoder and modulator chirps can be chosen independently. For a single user with zero-offset localized allocation, we show that these two chirps merge into a single residual chirp whose rate, with the other two chirp rates fixed, alone determines their effect. At the transmitter, this rate defines a circulant spreading kernel whose fourth moment quantifies the envelope fluctuation; symmetries reduce its characterization to rates in $[0,1/4]$. At a linear minimum mean-square error (\glsentryshort{mmse}) receiver, the residual chirp redistributes the mean-square error across symbols without changing its total; under a Gaussian approximation, a more uniform per-symbol error lowers the uncoded bit error rate (\glsentryshort{ber}). In uncoded \glsentryshort{qpsk} simulations, a zero rate yields the lowest \glsentryshort{papr}, whereas a generic rate reduces the signal-to-noise ratio required for a \glsentryshort{ber} of $10^{-3}$ by 4.4\,dB at a 3.33\,dB \glsentryshort{papr} cost (exceedance probability $10^{-3}$); a rate of $1/3$ captures 2.8\,dB of this gain for only 0.88\,dB of that cost.
\end{abstract}

\begin{IEEEkeywords}
\glsentryshort{afdm}, \glsentryshort{daft} spreading, high-mobility uplink, peak-to-average power ratio, bit error rate, linear \glsentryshort{mmse} equalization, doubly dispersive channels, \glsentryshort{sixg}.
\end{IEEEkeywords}

\section{Introduction}
\label{sec:intro}

In high-mobility uplink scenarios, such as vehicle-to-everything and high-speed railway communications~\cite{noorarahim2022v2x,ai2014hsr}, two challenges must be addressed simultaneously: reliable transmission over a doubly dispersive channel, where the Doppler spread destroys the orthogonality of the \gls{ofdm} subcarriers, and a well-behaved transmit envelope for devices with power amplifier constraints. For the former, \gls{afdm} is a promising solution: its configurable chirps impose a structured representation of the doubly dispersive channel that enables full-diversity detection~\cite{bemani2023afdm,rou2024otfs2afdm}.

The latter is addressed in \gls{fiveg} \gls{nr} through \gls{dftsofdm}\glsunset{dft}\glsunset{ofdm}, which addresses this issue through a \gls{dft} precoding stage \cite{myung2006scfdma,3gpp38211}.
A natural approach is therefore to combine both ideas, that is, the \gls{daftsafdm}\glsunset{daft}\glsunset{afdm}, in which a \gls{daft} precoder  is placed before the \gls{afdm} modulator, with the objective of retaining the delay--Doppler robustness of \gls{afdm} while obtaining the envelope benefits of precoding.
Standardization-oriented work has outlined how \gls{afdm} could be carried within
a \gls{dftsofdm} uplink~\cite{luo2026std}.
This contributes to a possible adoption of \gls{afdm} in \gls{imt}-2030 systems \cite{itu2160,rou2026afdm6g}, since \gls{daftsafdm} can fulfill the role \gls{dftsofdm} has in \gls{fiveg} \gls{nr}.

Unlike conventional \gls{dftsofdm}, \gls{daftsafdm} contains configurable chirp parameters in both the precoding and modulation stages. The relation between these chirp rates, therefore, becomes an additional design degree of freedom.
Tao et al. \cite{tao2024globecom} introduced \gls{daft}-spread \gls{afdma} for downlink transmission and derived chirp-parameter relations for localized and interleaved allocation. Their subsequent uplink study showed an important tradeoff: equal chirp rates on the precoder and the modulator lead to reduced \gls{papr}, while unequal chirp rates improved the \gls{ber} in the evaluated scenarios \cite{tao2026daftsafdma}. \Gls{daftsafdm} has also been explored in the context of \gls{isac}, including ambiguity-function analysis and constellation shaping \cite{cui_isac}.

Chirp parameter optimization has also been studied for conventional \gls{afdm}. \Gls{ber}-oriented chirp-selection rules under \gls{mmse} equalization were derived in \cite{li2025chirpmmse}, while joint chirp selection with reduced complexity \gls{mmse} reception was studied in \cite{mao2026jointchirp}. In parallel, pre-chirp selection and chirp tuning have been used to reduce the \gls{papr} of \gls{afdm} \cite{yuan2025,savaux2026}. These works show that the chirp rates affect both the transmitter envelope and receiver performance. Other spread or precoded \gls{afdm} variants target \gls{papr} reduction directly~\cite{lu2026a2fdm,ali2026spreading}, and chirped \gls{dftsofdm} offers a related single-carrier alternative~\cite{liu2025chirped}.

However, the joint effect of the precoder and modulator chirp rates in \gls{daftsafdm} has been studied only for a few specific combinations \cite{tao2024globecom,tao2026daftsafdma}. Those combinations represent only the extremes of the design space, with one side favoring low \gls{papr} and the other side favoring low \gls{ber}. They also do not explain what happens between these two cases or the mechanism behind this tradeoff. 
Motivated by this, we analyze the effect of the residual chirp rate between the precoding \gls{daft} and the modulating \gls{idaft} for a single-user \gls{daftsafdm} uplink with localized allocation and a linear \gls{mmse} receiver. Varying this rate yields a configurable \gls{papr}--\gls{ber} tradeoff between a \gls{dftsofdm}-like and an \gls{afdm}-like operating point.
Intermediate rates provide operating points between these two extremes, allowing the balance between \gls{papr} and \gls{ber} to be adjusted according to the system requirements.

The main contributions of this paper are as follows:
\begin{itemize}
\item For a single user with zero-offset localized allocation, we show that the
chirp rates at the precoder output and at the modulator input affect the
transmit signal only through their difference, the residual chirp rate, which
acts on the precoded symbols as a unitary circulant kernel. For an even number
of precoded symbols, symmetries of this kernel reduce the characterization of
its effective spread to residual chirp rates in $[0,1/4]$, although this does
not reduce the \gls{papr}--\gls{ber} design space in the same way.
\item At the transmitter, the effective spread of the kernel sets the fourth
moment of the spread symbols, which tracks the bulk of the \gls{papr}
distribution. In the linear \gls{mmse} receiver, the presence of the residual chirp is represented via unitary similarity, which does not change the total \gls{mse}, but redistributes it between the symbols, and the input chirp of the precoder does not change the per-symbol \gls{mse}. Under a Gaussian approximation, a more uniform per-symbol \gls{mse} lowers the uncoded \gls{ber}.
\item Simulations over a doubly dispersive channel map the \gls{papr}--\gls{ber}
operating points against \gls{dftsofdm} and \gls{afdm}. A zero residual chirp
rate gives the \gls{papr} of \gls{dftsofdm} at about 2\,dB lower required
\gls{snr}, generic rates perform like \gls{afdm}, and small-denominator rates
such as $1/3$ provide intermediate operating points.
\end{itemize}

 The remainder of this paper is organized as follows. Section~\ref{sec:system_model} presents the \gls{daftsafdm} transceiver, the doubly dispersive channel, and the residual-phase reduction. The effect of the residual chirp rate on spreading, \gls{papr}, and \gls{mmse} detection is analyzed in Section~\ref{sec:residualchirp}. Section~\ref{sec:results} presents the numerical \gls{papr}--\gls{ber} tradeoff, and Section~\ref{sec:conclusion} concludes the paper.

 \textit{Notations:}~ The following notations are adopted throughout this manuscript. Bold lowercase and uppercase letters denote vectors and matrices. The operators $(\cdot)^H$, $(\cdot)^T$, $\operatorname{diag}(\cdot)$, and $\mathbf I_L$ denote the Hermitian transpose, the ordinary transpose, a diagonal matrix, and the $L\times L$ identity matrix, respectively. The operators $\operatorname{tr}(\cdot)$, $\mathbb E[\cdot]$, and $\gcd(\cdot,\cdot)$ denote the trace, expectation, and greatest common divisor. Moreover, $j=\sqrt{-1}$, $\triangleq$ denotes a definition, $\bmod$ is the modulo operation, $\mathcal{CN}(\boldsymbol\mu,\boldsymbol\Sigma)$ denotes a circularly symmetric complex Gaussian distribution, and $\|\cdot\|$ denotes the Euclidean norm. Finally, $a\mid b$ means that the integer $a$ divides $b$. Unless stated otherwise, indices start at zero.

\section{System Model}
\label{sec:system_model}

\begin{figure}[t]
\centering
\input{figures/system_model_col.tikz}
\caption{Single-user \gls{daftsafdm} uplink.}
\label{fig:system_model}
\end{figure}
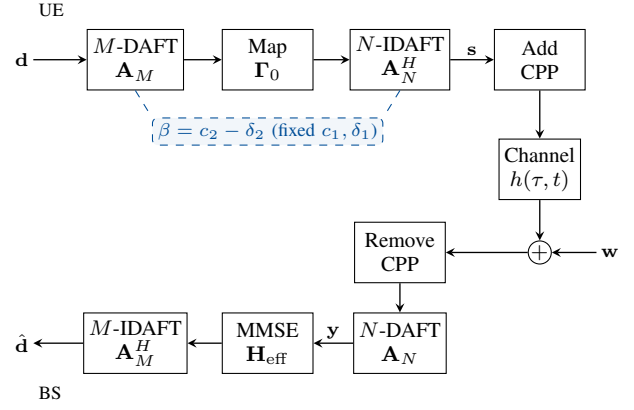

The system model for \gls{daftsafdm} is shown in Fig.~\ref{fig:system_model}. In the considered system, a single user transmits $M$ independent, equiprobable, unit-energy uncoded \gls{qpsk} symbols, expressed by $\mathbf d$. We assume perfect synchronization, perfect channel knowledge at the receiver, and a linear transmitter. The $L$-point \gls{daft} is defined as

\begin{equation}
\mathbf{A}_L(a_1,a_2) \triangleq \mathbf{\Lambda}_L(a_2)\,\mathbf{F}_L\,\mathbf{\Lambda}_L(a_1),
\label{eq:daft_matrix}
\end{equation}
where $\mathbf{F}_L$ the unitary \gls{dft} matrix with entries $[\mathbf F_L]_{k,r}=L^{-1/2}e^{-j2\pi kr/L}$ and $\mathbf{\Lambda}_L(c)$ is the diagonal chirp matrix, defined as
\begin{equation}
\mathbf{\Lambda}_L(c) \triangleq \operatorname{diag}\left( e^{-j2\pi c r^2} \right)_{r=0}^{L-1}
\label{eq:chirp_diag}
\end{equation}

The modulator is composed of a $M$-point \gls{daft} spreading \cite{tao2026daftsafdma,cui_isac} matrix $\mathbf A_M$ followed by an inverse \gls{daft} modulator matrix $\mathbf A_N^H$ (where  $N>M$), while the mapping of the matrices of size $M$ to $N$ is done through the $\boldsymbol\Gamma$ in between the precoding and modulation matrices. Since in this paper we only consider the localized mapping strategy for a single user, $\boldsymbol\Gamma_0=[\mathbf I_M\;\mathbf 0]^T\in\{0,1\}^{N\times M}$. The full transmitter can be written as:
\begin{equation}
\mathbf{s} = \mathbf{A}_N^H\,\boldsymbol\Gamma_0\,\mathbf A_M\,\mathbf{d}
\label{eq:tx_vector}
\end{equation}

Expanding the \gls{daft} and \gls{idaft} factors in \eqref{eq:tx_vector} leads to
\begin{equation}
\begin{split}
\mathbf{s}={}&\underbrace{\mathbf\Lambda_N(c_1)^H\mathbf F_N^H\mathbf\Lambda_N(c_2)^H}_{N\text{-point \glsentryshort{idaft} modulator }\mathbf A_N^H}\,\boldsymbol\Gamma_0\\
&\times\underbrace{\mathbf\Lambda_M(\delta_2)\mathbf F_M\mathbf\Lambda_M(\delta_1)}_{M\text{-point \glsentryshort{daft} precoder }\mathbf A_M}\,\mathbf{d}.
\end{split}
\label{eq:tx_vector_full}
\end{equation}

In \eqref{eq:tx_vector_full}, the four chirp rates play distinct roles: $c_1$ separates delay--Doppler paths at the modulator, and $\delta_1$ sets the precoder phase progression. The other two are not independent in the considered scenario: since the localized mapping places the precoder output in the first $M$ bins, $\mathbf\Lambda_N(c_2)^H\boldsymbol\Gamma_0=\boldsymbol\Gamma_0\mathbf\Lambda_M(c_2)^H$, so the two chirp diagonals become adjacent and their product simplifies to
\begin{equation}
\begin{split}
\mathbf\Lambda_M(c_2)^H\mathbf\Lambda_M(\delta_2) &=\operatorname{diag}\!\left(e^{j2\pi(c_2-\delta_2) m^2}\right)_{m=0}^{M-1} \\
&=\mathbf\Lambda_M(c_2-\delta_2)^H
\end{split}
\label{eq:residual_phase}
\end{equation}

The residual chirp rate parameter is defined as $\beta\triangleq c_2-\delta_2$. It is then possible to rewrite \eqref{eq:tx_vector} in compact form as:
\begin{equation}
\mathbf s=\mathbf Q\boldsymbol\Gamma_0\mathbf U\mathbf d,
\label{eq:tx_reduced}
\end{equation}
where $\mathbf U=\mathbf A_M(\delta_1,-\beta)$ and $\mathbf Q\triangleq\mathbf A_N(c_1,0)^H=\mathbf\Lambda_N(c_1)^H\mathbf F_N^H$.
For an allocation starting at bin $k_0\neq0$, $c_2$ also contributes a linear phase, which we leave to future work.

Equation~\eqref{eq:tx_reduced} isolates the role of the residual chirp rate. In the matched case, $\beta=0$ (i.e., $c_2=\delta_2$), the adjacent chirp matrices cancel and $\mathbf U=\mathbf F_M\mathbf\Lambda_M(\delta_1)$. Hence, $\beta$ is the parameter that controls the change from the matched baseline, and its effects on \gls{papr} and \gls{ber} are analyzed in Section~\ref{sec:residualchirp}.
The transmit block is then localized \gls{dftsofdm} with symbols rotated by
$\mathbf\Lambda_M(\delta_1)$, multiplied by the unit-modulus chirp
$\mathbf\Lambda_N(c_1)^H$; its envelope is therefore that of \gls{dftsofdm}, while
the chirp sweeps the occupied band across the block.

\subsection{Doubly Dispersive Uplink Channel}
\label{subsec:dd_channel}

The doubly dispersive channel model used throughout this paper is defined as
\begin{equation}
h(\tau, t) = \sum_{p=1}^{P} h_{p}\, \delta(\tau - \tau_{p})\, e^{j 2\pi \nu_{p} t},
\label{eq:cir}
\end{equation}
where $P$ is the number of resolvable paths and $h_{p}$, $\tau_{p}$, and $\nu_{p}$ denote the complex gain, delay, and Doppler shift of the $p$-th path. It is assumed that its gains, delays and Doppler does not change over one block period.

Let $T_s$ denote the sample period. We assume integer sample delays $l_p=\tau_p/T_s$, with $0\le l_p\le l_{\max}<N$, and define $f_p=\nu_pNT_s$, which may be fractional. 
Like in standard \gls{afdm}, the signal is appended with a chirp-periodic prefix (\glsentryshort{cpp})\glsunset{cpp} of length $L_{\mathrm{CPP}}\ge l_{\max}$ at the transmitter and removed at the receiver. The resulting circular channel matrix $\mathbf{H}$ is given as
\begin{equation}
\mathbf{H} = \sum_{p=1}^{P} h_{p}\, \mathbf{D}_{\mathrm{CPP},p}\, \boldsymbol\Delta_{p}\, \boldsymbol\Pi^{\,l_{p}},
\label{eq:channel_matrix}
\end{equation}
where $\boldsymbol\Pi^{\,l_{p}}$ is a circular shift matrix, $\boldsymbol\Delta_{p}=\operatorname{diag}(e^{j2\pi f_{p} n/N})_{n=0}^{N-1}$ is the Doppler diagonal matrix, and the diagonal matrix $\mathbf D_{\mathrm{CPP},p}$ applies the phase that the \gls{cpp} imposes on the $l_p$ wrapped-around samples~\cite{bemani2023afdm}. This phase equals one whenever $2Nc_1$ and $c_1N^2$ are integers, as in our simulations; then $\mathbf D_{\mathrm{CPP},p}=\mathbf I_N$ and the \gls{cpp} reduces to a cyclic prefix.

For integer Doppler indices bounded by $\alpha_{\max}$ and delays bounded by
$l_{\max}$, distinct delay--Doppler paths are separated when
$2Nc_1\ge2\alpha_{\max}+1$ and $l_{\max}\,2Nc_1+2\alpha_{\max}+1\le N$, the
minimal choice being $c_1=(2\alpha_{\max}+1)/(2N)$~\cite{bemani2023afdm}; a guard
of $k_g$ bins replaces $\alpha_{\max}$ by $\alpha_{\max}+k_g$ to leave margin for
the leakage of fractional Doppler.

\subsection{Affine-Domain \glsentryshort{mmse} Receiver}

After \gls{cpp} removal and the receiver \gls{daft}, the received signal can be expressed as
\begin{equation}
    \mathbf y=\mathbf H_{\mathrm{eff}}\mathbf A_M\mathbf d+\mathbf A_N\mathbf w,
    \label{eq:received_signal}
\end{equation}
where $\mathbf H_{\mathrm{eff}}=\mathbf A_N\mathbf H\mathbf A_N^H\boldsymbol\Gamma_0$ and \gls{awgn} $\mathbf w\sim\mathcal{CN}(\mathbf0,N_0\mathbf I_N)$. The receiver equalizes the precoded symbols using $\mathbf H_{\mathrm{eff}}$ and then deprecodes:
\begin{equation}
\hat{\mathbf d}=\mathbf A_M^H\big(\mathbf H_{\mathrm{eff}}^H\mathbf H_{\mathrm{eff}}
+\gamma^{-1}\mathbf I_M\big)^{-1}\mathbf H_{\mathrm{eff}}^H\mathbf y,
\label{eq:mmse_filter}
\end{equation}
where $\gamma=E_s/N_0=1/N_0$, since the \gls{qpsk} symbols have unit energy. As
the block carries $M$ unit-energy symbols over $N$ samples, the \gls{snr} per
useful sample is $(M/N)\gamma$, i.e., 3\,dB below $\gamma$ for $M=N/2$.

Let $\boldsymbol\varepsilon=\mathbf d-\hat{\mathbf d}$ denote the estimation error at the output of \eqref{eq:mmse_filter}. Conditioned on the channel, its covariance is
\begin{equation}
\begin{split}
\mathbf C_e
&\triangleq\mathbb E\!\left[
\boldsymbol\varepsilon\boldsymbol\varepsilon^H\mid\mathbf H
\right]\\
&=\mathbf A_M^H
\left(\mathbf I_M+\gamma\mathbf H_{\mathrm{eff}}^H
\mathbf H_{\mathrm{eff}}\right)^{-1}\mathbf A_M\\
&=\mathbf U^H\mathbf C_0\mathbf U,
\end{split}
\label{eq:error_cov}
\end{equation}
where
\begin{equation}
\mathbf C_0=
\left(\mathbf I_M+\gamma\boldsymbol\Gamma_0^H\mathbf Q^H
\mathbf H^H\mathbf H\mathbf Q\boldsymbol\Gamma_0\right)^{-1}.
\label{eq:ref_cov}
\end{equation}
For fixed $\mathbf H$, $\gamma$, $c_1$, and allocation, $\mathbf C_0$ is independent of $\beta$ and $\delta_1$. Since $\mathbf U$ is unitary, \eqref{eq:error_cov} is a unitary similarity and therefore
\begin{equation}
\operatorname{tr}(\mathbf C_e)=\operatorname{tr}(\mathbf C_0).
\label{eq:total_mse}
\end{equation}
Thus, $\beta$ preserves the total block \gls{mse} but can change the individual errors $e_i=[\mathbf C_e]_{i,i}$ and the off-diagonal error correlations.

\section{Residual Chirp Analysis}
\label{sec:residualchirp}

Both metrics depend on $\beta$ through a single matrix.
\begin{proposition}\label{prop:kernel}
Let
\begin{equation}
\boldsymbol\Xi_\beta\triangleq\mathbf F_M^H\mathbf\Lambda_M(\beta)^H\mathbf F_M .
\label{eq:mixing_matrix}
\end{equation}
Then $\mathbf U=\mathbf F_M\boldsymbol\Xi_\beta\mathbf\Lambda_M(\delta_1)$, so that
\begin{align}
\mathbf s&=\mathbf Q\boldsymbol\Gamma_0\mathbf F_M\,\boldsymbol\Xi_\beta\,
\mathbf\Lambda_M(\delta_1)\,\mathbf d,
\label{eq:mixing_kernel}\\
\mathbf C_e&=\mathbf\Lambda_M(\delta_1)^H\boldsymbol\Xi_\beta^H\,
\mathbf C_e^{(0)}\,\boldsymbol\Xi_\beta\mathbf\Lambda_M(\delta_1),
\quad \mathbf C_e^{(0)}\triangleq\mathbf F_M^H\mathbf C_0\mathbf F_M,
\label{eq:ce_kernel}
\end{align}
where $\mathbf C_e^{(0)}$ is the error covariance for $\beta=\delta_1=0$.
\end{proposition}
The proof follows from $\mathbf F_M\boldsymbol\Xi_\beta=\mathbf\Lambda_M(\beta)^H\mathbf F_M$
and $\mathbf U=\mathbf A_M(\delta_1,-\beta)=\mathbf\Lambda_M(\beta)^H\mathbf F_M\mathbf\Lambda_M(\delta_1)$,
substituted into \eqref{eq:tx_reduced} and \eqref{eq:error_cov}.
Being unitary, $\boldsymbol\Xi_\beta$ conserves second-order quantities at both
ends: the total energy of its taps is one, which fixes the mean transmit power,
and $\operatorname{tr}(\mathbf C_e)=\operatorname{tr}(\mathbf C_0)$ fixes the total
\gls{mse}. The residual chirp rate therefore only redistributes: at the
transmitter, each spread symbol combines several data symbols; at the receiver,
each symbol's error combines the $\beta=0$ errors at several time positions.
Sections~\ref{subseq:PAPR} and~\ref{subsec:BER} show that this same
redistribution raises the \gls{papr} and lowers the \gls{ber}.

\subsection{Effects on \glsentryshort{papr}}
\label{subseq:PAPR}
For a single transmitted block, the sampled \gls{papr} is defined as
\begin{equation}
\mathrm{PAPR}\triangleq\frac{\max_n|s[n]|^2}{N^{-1}\sum_n|s[n]|^2}.
\label{eq:papr_def}
\end{equation}
Since $\mathbf Q\boldsymbol\Gamma_0\mathbf U$ has orthonormal columns and
$\|\mathbf d\|^2=M$ for \gls{qpsk}, the denominator equals $M/N$ for every block,
and removing the unit-modulus chirp $\mathbf\Lambda_N(c_1)^H$ leaves all sample
powers unchanged. We denote by $\mathrm{PAPR}_J$ the same ratio evaluated on
$J\ge N$ uniform samples of the Fourier interpolation of the dechirped block;
Section~\ref{sec:results} uses $J=4N$. This reconstruction corresponds to an analog-chirp implementation; digital application of the chirp before interpolation can increase the peak values, while other pulses need individual attention, as analyzed in the continuous-time case in~\cite{mirabella2026continuous}.

The matrix $\boldsymbol\Xi_\beta$ in \eqref{eq:mixing_matrix} can be seen as a generalized discrete Fresnel transformation (\glsentryshort{dfnt})\glsunset{dfnt} matrix with a free parameter $\beta$~\cite{OuyangOCDM_TCOM2016}.\footnote{The $M$-point \gls{dfnt} is defined for even $M$ in~\cite{OuyangOCDM_TCOM2016} as the unitary circulant matrix $\boldsymbol\Phi$ with entries $[\boldsymbol\Phi]_{m,n}=M^{-1/2}e^{-j\pi/4}e^{j\pi(m-n)^2/M}$, whose eigenvalues form a root Zadoff--Chu sequence, i.e., $\boldsymbol\Phi=\mathbf F_M^H\mathbf\Lambda_M\!\left(\tfrac{1}{2M}\right)\mathbf F_M$. Comparing this with \eqref{eq:mixing_matrix} gives $\boldsymbol\Xi_{-1/(2M)}=\boldsymbol\Phi$ and $\boldsymbol\Xi_{1/(2M)}=\boldsymbol\Phi^H$. Hence $\boldsymbol\Xi_\beta$ generalizes the \gls{dfnt}, a discrete analogue of the fractional Talbot effect~\cite{OuyangOCDM_TCOM2016}, to an arbitrary chirp rate $\beta$ with period one. All entries of $\boldsymbol\Phi$ have magnitude $M^{-1/2}$, so the \gls{dfnt} itself spreads each symbol over all $M$ taps ($S_{\mathrm{eff}}=M$).}
Any matrix diagonalized by $\mathbf F_M$ is circulant, so \eqref{eq:mixing_matrix} acts as a circular filter across the precoded symbols. Thus, the taps of $\boldsymbol\Xi_\beta$ can be written as
\begin{equation}
\xi_\beta[\ell]=\frac1M\sum_{m=0}^{M-1}e^{j2\pi(\beta m^2+m\ell/M)},\qquad 0\le\ell<M .
\label{eq:kernel_taps}
\end{equation}

With $\mathbf a=\mathbf\Lambda_M(\delta_1)\mathbf d$, the spread symbols $\mathbf v=\boldsymbol\Xi_\beta\mathbf a$ are $v[r]=\sum_{\ell}\xi_\beta[\ell]\,a[(r-\ell)\bmod M]$: each is a weighted sum of circularly shifted symbols, with constant magnitude only if the kernel has a single tap.

Since $\boldsymbol\Xi_\beta$ is unitary, the total energy of its taps is one. Hence, $p_\beta[\ell]\triangleq|\xi_\beta[\ell]|^2$ defines a normalized distribution of the kernel energy over the circular shifts $\ell$. We then define $S_{\mathrm{eff}}$ as the inverse participation ratio of this distribution as
\begin{equation}
S_{\mathrm{eff}}(\beta)
=\left(\sum_{\ell=0}^{M-1}|\xi_\beta[\ell]|^4\right)^{-1},
\qquad 1\le S_{\mathrm{eff}}\le M.
\label{eq:effective_spread}
\end{equation}

For independent, proper ($\mathbb E\{a[r]^2\}=0$), unit-energy symbols with $\kappa=\mathbb E\{|a[r]|^4\}$, the spread symbols satisfy $\mathbb E\{|v[r]|^4\}=2-(2-\kappa)/S_{\mathrm{eff}}(\beta)$, which for \gls{qpsk} ($\kappa=1$) gives
\begin{equation}
\mathbb E\{|v[r]|^4\}
=2-\frac{1}{S_{\mathrm{eff}}(\beta)}.
\label{eq:fourth_moment}
\end{equation}

Several values of $\beta$ generate kernels whose taps differ only by conjugation,
reversal, or circular shift, which preserve $S_{\mathrm{eff}}$. Since $m^2$ is an
integer, $\boldsymbol\Xi_{\beta+1}=\boldsymbol\Xi_\beta$ and
$\boldsymbol\Xi_{1-\beta}=\boldsymbol\Xi_\beta^H$; for even $M$, $m^2$ and $m$
have the same parity, so $e^{j\pi m^2}=(-1)^m$ and $\boldsymbol\Xi_{\beta+1/2}$
is $\boldsymbol\Xi_\beta$ combined with a circular shift by $M/2$. Hence
$S_{\mathrm{eff}}(\beta)=S_{\mathrm{eff}}(1-\beta)=S_{\mathrm{eff}}(\beta+1/2)$,
which reduces the characterization of $S_{\mathrm{eff}}$ to $\beta\in[0,1/4]$.
The half-period shift also permutes the per-symbol \glspl{mse} of
Section~\ref{subsec:BER} in every realization and, since $4M\delta_1$ is an
integer here, leaves the \gls{papr} distribution unchanged.

The preceding identities reduce the $\beta$ values interval but do not yet quantify the spread at a given point. Spreading tap count is available when the residual chirp rate is rational. Let $\beta=u/q$ and assume that $q\mid M$. In this case, the block contains an integer number of periods of the quadratic-phase sequence, and the Gauss sum vanishes at all but $S$ equally spaced kernel positions, where the taps have equal magnitude. Their number is
\begin{equation}
S=\frac{q}{\gcd(q,2)}
\label{eq:rational_taps}
\end{equation}
and therefore $S_{\mathrm{eff}}=S$. Then its possible to determine the number of taps, an odd $q$ gives $q$ taps, whereas an even $q$ gives $q/2$ taps. 

When $q\nmid M$, the block does not contain the required integer number of phase periods, the cancellations are incomplete and the taps are no longer exactly equal. In that case, $S_{\mathrm{eff}}$ must be evaluated directly from \eqref{eq:effective_spread}. Yet, $\beta$ as irrational does not imply maximum spreading, since an irrational value sufficiently close to zero still produces a kernel close to the one-tap case. Therefore, the spread, and consequently the general \gls{papr} trend, is determined by the block kernel rather than by the numerical size or rationality of $\beta$ alone \cite{ochiai2001papr}.

\subsection{Effects on \glsentryshort{ber}}
\label{subsec:BER}

By \eqref{eq:ce_kernel}, the outer diagonal unitary matrices leave
$\operatorname{diag}(\mathbf C_e)$ unchanged, so the per-symbol \glspl{mse},
\glspl{sinr}, and \gls{ber} depend on $\beta$ but not on $\delta_1$.

Proposition~\ref{prop:kernel} also shows how $\beta$ evens out the per-symbol \gls{mse}. At the transmit instants $n=rN/M$, $|s[n]|=\sqrt{M/N}\,|v[r]|$, so the index of a spread symbol is a time position. At $\beta=0$, each data symbol occupies its own time slot; under Doppler, the channel gain varies across the block, and symbols falling into a fade suffer a large $e_i$. For $\beta\neq0$, the kernel places copies of each symbol at the positions of its nonzero taps; for $\beta=u/q$ with $q\mid M$, these are $S$ equal-weight copies spaced $N/S$ samples apart. The diagonal of \eqref{eq:ce_kernel} then gives
\begin{equation}
e_i(\beta)=\sum_{k}|\xi_\beta[(k-i)\bmod M]|^2\,e_k(0)+r_{\beta,i},
\label{eq:mse_average}
\end{equation}
where the first term is a weighted average of the $\beta=0$ errors at the copy positions, which is never less uniform than the $\beta=0$ errors themselves, and $r_{\beta,i}$ collects the correlations between those errors. The correlation term sums to zero, while the averaging term explains the tendency toward a more uniform error.

The residual interference and filtered noise in the $i$-th entry of $\hat{\mathbf d}$ have total
variance $e_i(1-e_i)$, so the unbiased post-equalization \gls{sinr}
is~\cite{verdu1998mud}
\begin{equation}
\mathrm{SINR}_i(\beta)=\frac{1-e_i}{e_i},
\label{eq:post_sinr}
\end{equation}
and, under a proper-complex-Gaussian approximation for the residual interference, the \gls{qpsk} \gls{ber} is 
\begin{equation}
\mathrm{BER}_i(\beta)\approx
Q_{\mathrm G}\!\left(\sqrt{\mathrm{SINR}_i(\beta)}\right).
\label{eq:qpsk_ber_proxy}
\end{equation}
By \eqref{eq:total_mse}, $\beta$ cannot change the total error
$\sum_ie_i=\operatorname{tr}(\mathbf C_0)$; it can only move error between symbols.
The function $g(e)\triangleq\mathrm Q_{\mathrm G}\big(\sqrt{(1-e)/e}\big)$ is
increasing and convex on $(0,1)$, since its second derivative is a positive
multiple of $(1-2e)^2$. The average \gls{ber} $M^{-1}\sum_i g(e_i)$ is therefore
Schur-convex in $(e_0,\dots,e_{M-1})$~\cite{palomar2003}: among per-symbol
\gls{mse} vectors with the same total, a more uniform vector, in the
majorization sense, yields a lower \gls{ber}, and equal $e_i$ yield the lowest.
The residual chirp rate can thus lower the uncoded \gls{ber} without reducing the
total error, by distributing it more evenly across the symbols.

Both metrics thus respond to the same redistribution, in opposite directions:
spreading makes each transmit sample a sum of about $S_{\mathrm{eff}}$ symbols,
so the envelope fluctuates more, while it makes each symbol's error an average
over about $S_{\mathrm{eff}}$ time positions, so the error fluctuates less. At
$\beta=0$ nothing is redistributed (lowest \gls{papr}, most uneven error); generic
$\beta$ redistributes almost fully (near-Gaussian envelope, near-uniform error);
and only small-denominator $\beta$, whose $S_{\mathrm{eff}}$ does not grow with
$M$, yield intermediate operating points. Section~\ref{sec:results} quantifies
this tradeoff.

\section{Simulation Results}
\label{sec:results}

We simulate $M=64$ \gls{qpsk} symbols mapped to the first $M$ of $N=128$ bins, over $P=3$ paths with Rayleigh gains normalized to unit total power in each realization, integer delays up to $l_{\max}=3$ samples, and normalized Doppler shifts following the Jakes model $f_p=\alpha_{\max}\cos\theta_p$ with
$\alpha_{\max}=3$ and $\theta_p$ uniform, as in~\cite{tao2026daftsafdma}. With $f_c=50$~GHz and a subcarrier spacing of $7.8$~kHz ($f_s=1$~MHz), this corresponds to a maximum speed of about 500~km/h and a delay spread of 3~$\mu$s. The modulator uses $c_1=15/128$ with a guard of $k_g=4$ bins, and the precoder uses $\delta_1=15/64$, which plays no path-separation role and leaves the per-symbol \gls{mse} unchanged.
The \gls{ber} is averaged over 1000 independent channel,
noise, and data realizations per \gls{snr} point, the paths have
equal average power before the normalization, and
$L_{\mathrm{CPP}}=[l_{\max}]$.
For the comparison, all schemes use the same $M$ unit-energy \gls{qpsk} symbols,
the same $N$ useful samples, the same localized mapping, and the same channel and
noise realizations; the useful block energy and the prefix length are also the
same. \gls{daftsafdm} and \gls{afdm} use the modulation and prefix of
Section~\ref{sec:system_model}, whereas \gls{dftsofdm} uses \gls{dft} precoding,
the same localized mapping, an $N$-point \gls{idft}, and a conventional cyclic
prefix. In all cases, the receiver has perfect channel knowledge and applies
linear \gls{mmse} equalization to the corresponding effective channel.
The evaluated residual chirp rates are $\beta\in\{0,1/3,1/16\}$, with
$S_{\mathrm{eff}}=1$, $4.49$, and $8$ at $M=64$, together with the generic
value $\beta=\pi\bmod1$, for which $S_{\mathrm{eff}}=52.6$.

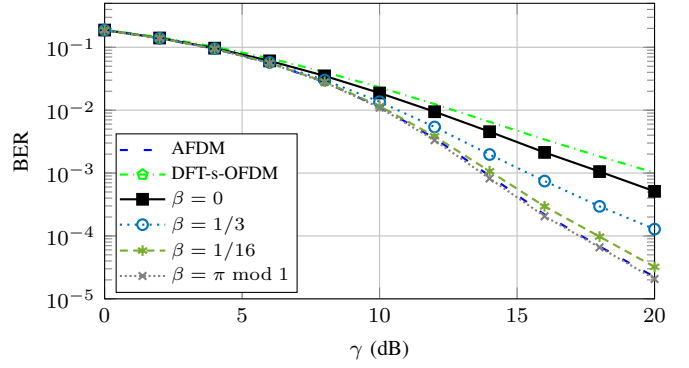
\begin{figure}[t]
\centering
\input{figures/ber_beta_grid_conf_v13.tikz}
\caption{Simulated \glsentryshort{ber} versus $\gamma$}
\label{fig:ber_beta_grid}
\end{figure}

\begin{figure}[t]
\centering
\input{figures/papr_beta_grid_conf_v13.tikz}
\caption{Simulated \glsentryshort{ccdf} of $\mathrm{PAPR}_{4N}$. Legend as in Fig.~\ref{fig:ber_beta_grid}.}
\label{fig:papr_beta_grid}
\end{figure}
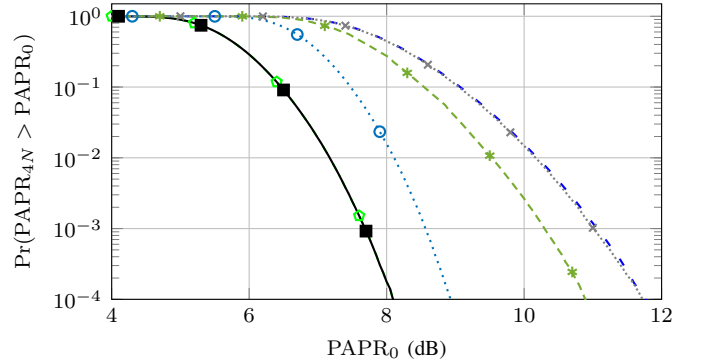

\subsection{\glsentryshort{ber} Results}

Fig.~\ref{fig:ber_beta_grid} shows the resulting \gls{ber}.
At a \gls{ber} of $10^{-3}$, matching the two chirp rates ($\beta=0$) provides a
gain of approximately 2\,dB over \gls{dftsofdm} while remaining in the
low-\gls{papr} operating region. Relative to $\beta=0$, the residual chirp rates
$\beta=1/3$ and $1/16$ reduce the required \gls{snr} by a further 2.8 and
4.0\,dB, respectively, and $\beta=\pi\bmod1$ by 4.4\,dB, practically overlapping
\gls{afdm}. The gain grows with the kernel spread $S_{\mathrm{eff}}$, and saturates beyond $S_{\mathrm{eff}}\approx8$:
$\beta=1/16$ is within 0.4\,dB of the generic value, although its kernel has far
fewer taps.

This ordering is consistent with the covariance mechanism of Section~\ref{subsec:BER}. A nonzero $\beta$ does not change the total \gls{mse}, but it redistributes this error more uniformly among the detected symbols.
These results explain the behavior reported in~\cite{tao2026daftsafdma}, whose
precoder chirp $\lambda_2$ corresponds to our $\delta_2$: the equal setting
$\lambda_2=c_2$ gives $\beta=0$, while the unequal settings
$\lambda_2\in\{0,0.25\pi,0.99\pi\}$ with $c_2=\pi$ give generic values
($\beta\bmod1\approx0.14$, $0.36$, $0.03$) with $S_{\mathrm{eff}}$ between 165
and 290 at $M=512$, hence AFDM-like \gls{papr}; closeness in $\beta$ is measured
on a $1/M^2$ scale. Intermediate operating points require small-denominator
$\beta$, which were not evaluated there.

\subsection{\glsentryshort{papr} Results}

Fig.~\ref{fig:papr_beta_grid} shows the \gls{ccdf} of the \gls{papr}, estimated
from $10^6$ random blocks. At a \gls{ccdf} of $10^{-3}$, the \gls{papr} for
$\beta=0$ is close to that of \gls{dftsofdm} (7.68\,dB), while $\beta=1/3$,
$1/16$, and $\pi\bmod1$ require 8.56, 10.30, and 11.01\,dB, and \gls{afdm}
requires 11.06\,dB; the complete range from the \gls{dftsofdm}-like to the
\gls{afdm}-like case thus costs about 3.38\,dB.
The fourth moment in \eqref{eq:fourth_moment}, which approaches the Gaussian value 2 as more symbols are combined, tracks the bulk of the \gls{papr} distribution: as $2-S_{\mathrm{eff}}^{-1}$ increases through 1, 1.78, 1.875, and 1.98, the median \gls{papr} increases through 5.7, 6.8, 7.55, and 7.9\,dB. Measured at a \gls{ber} of $10^{-3}$ and a \gls{ccdf} of $10^{-3}$, $\beta=1/3$ attains 2.8\,dB of the 4.4\,dB \gls{snr} gain of the generic value for 0.88\,dB of its 3.33\,dB \gls{papr} cost, an efficient compromise, while $\beta=1/16$ comes within 0.4\,dB of the generic \gls{snr} gain at a 0.7\,dB lower \gls{papr}.

The value $\beta=1/4$ ($S_{\mathrm{eff}}=2$) is dominated by $\beta=1/3$ in \gls{ber} and in the \gls{papr} tail, and $\beta=1/6$ matches $\beta=1/3$ to within 0.05\,dB in both metrics, as the symmetries predict; both are omitted from the figures.

\section{Conclusion}
\label{sec:conclusion}

For a single user with zero-offset localized allocation, the precoder and modulator chirp rates of \gls{daftsafdm} affect the system only through their difference, the residual chirp rate, which acts as a unitary circulant kernel on the precoded symbols. The kernel conserves the transmit power and the total \gls{mse} but redistributes both: its effective spread raises the fourth moment of the spread symbols, which tracks the bulk of the \gls{papr} distribution, and makes the per-symbol \gls{mse} more uniform, which lowers the uncoded \gls{ber}. A zero residual chirp rate matches the \gls{papr} of \gls{dftsofdm}, generic values perform like \gls{afdm}, and a small-denominator value such as $1/3$ attains about 60\% of the \gls{snr} gain for about one quarter of the \gls{papr} cost. Future work includes considering coded transmission, higher-order constellations, multi-user allocation, imperfect channel knowledge, and interference-cancellation receivers.

\balance
\bibliographystyle{IEEEtran}
\bibliography{ref}

\end{document}

%% file: figures/system_model_col.tikz
\definecolor{betaacc}{RGB}{0,78,152}%
\begin{tikzpicture}[
  >=stealth, font=\footnotesize,
  block/.style={draw, rectangle, align=center, minimum height=0.8cm,
                minimum width=1.2cm, inner sep=2pt},
  flow/.style={->, line width=0.5pt},
  signal/.style={font=\scriptsize, inner sep=1.5pt}
]
\node[block] (prec) at (0.9,0) {$M$-DAFT\\$\mathbf A_M$};
\node[block] (map)  at (2.65,0) {Map\\$\boldsymbol\Gamma_0$};
\node[block] (mod)  at (4.4,0) {$N$-IDAFT\\$\mathbf A_N^H$};
\node[block] (cpp)  at (6.25,0) {Add\\CPP};
\draw[flow] (-0.45,0) node[left,signal]{$\mathbf d$} -- (prec);
\draw[flow] (prec) -- (map);
\draw[flow] (map) -- (mod);
\draw[flow] (mod) -- node[signal,above] {$\mathbf s$} (cpp);
\node[anchor=south west,font=\scriptsize] at (-0.5,0.45) {UE};
\node[draw=betaacc,dashed,rounded corners=2pt,text=betaacc,fill=betaacc!5,
      inner sep=2pt,font=\scriptsize] (beta) at (2.65,-0.95)
  {$\beta=c_2-\delta_2$ (fixed $c_1,\delta_1$)};
\draw[dashed,betaacc] (prec.south) -- (beta.north west);
\draw[dashed,betaacc] (mod.south) -- (beta.north east);
\node[block,minimum width=0.9cm] (chan) at (6.25,-1.45) {Channel\\$h(\tau,t)$};
\draw[flow] (cpp) -- (chan);
\node[draw,circle,inner sep=0pt,minimum size=3mm] (sum) at (6.25,-2.55) {$+$};
\draw[flow] (chan) -- (sum);
\draw[flow] (7.0,-2.55) node[right,signal]{$\mathbf w$} -- (sum);
\node[block] (rem)  at (4.4,-2.55) {Remove\\CPP};
\node[block] (daft) at (4.4,-3.75) {$N$-DAFT\\$\mathbf A_N$};
\node[block] (eq)   at (2.65,-3.75) {MMSE\\$\mathbf H_{\mathrm{eff}}$};
\node[block] (desp) at (0.9,-3.75) {$M$-IDAFT\\$\mathbf A_M^H$};
\draw[flow] (sum) -- (rem);
\draw[flow] (rem) -- (daft);
\draw[flow] (daft) -- node[signal,above] {$\mathbf y$} (eq);
\draw[flow] (eq) -- (desp);
\draw[flow] (desp) -- (-0.45,-3.75) node[left,signal] {$\hat{\mathbf d}$};
\node[anchor=north west,font=\scriptsize] at (-0.5,-4.2) {BS};
\end{tikzpicture}

%% file: figures/ber_beta_grid_conf_v13.tikz
\begin{tikzpicture}
\begin{semilogyaxis}[
    width=\linewidth,
    height=0.62\linewidth,
    xlabel={$\gamma$ (dB)},
    ylabel={BER},
    grid=major,
    xmin=0, xmax=20,
    ymin=1e-5, ymax=0.5,
    legend cell align={left},
    legend columns=1,
    legend style={at={(0.02,0.02)}, anchor=south west, font=\scriptsize,
                  inner sep=1pt, row sep=-1pt,
                  /tikz/every even column/.append style={column sep=4pt}},
    unbounded coords=jump
]

\definecolor{cB0}{RGB}{0,0,0}
\definecolor{cB13}{RGB}{0,114,189}
\definecolor{cB14}{RGB}{217,83,25}
\definecolor{cB16}{RGB}{126,47,142}
\definecolor{cB116}{RGB}{119,172,48}
\definecolor{cBgold}{RGB}{190,20,100}
\definecolor{cBpi}{RGB}{128,128,128}

\addplot[color=blue,    loosely dashed,          thick, mark=o*, mark repeat=24, mark phase=3,  mark options={solid}] table[x=SNR_dB, y=AFDM,        col sep=comma] {data/fig2_BER_data.csv};
\addplot[color=green,    dashdotted,          thick, mark=pentagon, mark repeat=24, mark phase=1,  mark options={solid}] table[x=SNR_dB, y=DFT_s_OFDM,        col sep=comma] {data/fig2_BER_data.csv};
\addplot[color=cB0,    solid,            thick, mark=square*, mark options={solid}] table[x=SNR_dB, y=beta_0,        col sep=comma] {data/fig2_BER_data.csv};
\addplot[color=cB13,   dotted,           thick, mark=o,        mark options={solid}] table[x=SNR_dB, y=beta_1_3,      col sep=comma] {data/fig2_BER_data.csv};
\addplot[color=cB116,  densely dashed,   thick, mark=asterisk, mark options={solid}] table[x=SNR_dB, y=beta_1_16,     col sep=comma] {data/fig2_BER_data.csv};
\addplot[color=cBpi,   densely dotted,   thick, mark=x,        mark options={solid}] table[x=SNR_dB, y=beta_pi,       col sep=comma] {data/fig2_BER_data.csv};

\legend{AFDM, $\text{DFT-s-OFDM}$, $\beta=0$, $\beta=1/3$, $\beta=1/16$, $\beta=\pi\bmod1$}
\end{semilogyaxis}
\end{tikzpicture}

%% file: figures/papr_beta_grid_conf_v13.tikz
\begin{tikzpicture}
\begin{semilogyaxis}[
    width=\linewidth,
    height=0.62\linewidth,
    xlabel={$\mathrm{PAPR}_0$ (dB)},
    ylabel={$\Pr(\mathrm{PAPR}_{4N}>\mathrm{PAPR}_0)$},
    xmin=4, xmax=12,
    ymin=1e-4, ymax=1.5,
    grid=major,
    unbounded coords=jump
]

\definecolor{cB0}{RGB}{0,0,0}
\definecolor{cB13}{RGB}{0,114,189}
\definecolor{cB14}{RGB}{217,83,25}
\definecolor{cB16}{RGB}{126,47,142}
\definecolor{cB116}{RGB}{119,172,48}
\definecolor{cBgold}{RGB}{190,20,100}
\definecolor{cBpi}{RGB}{128,128,128}

\addplot[color=blue,    loosely dashed,          thick, mark=o*, mark repeat=24, mark phase=3,  mark options={solid}] table[x=papr_dB, y=AFDM,        col sep=comma] {data/fig3_PAPR_data.csv};
\addplot[color=green,    dashdotted,          thick, mark=pentagon, mark repeat=24, mark phase=1,  mark options={solid}] table[x=papr_dB, y=DFT_s_OFDM,        col sep=comma] {data/fig3_PAPR_data.csv};

\addplot[color=cB0,    solid,          thick, mark=square*, mark repeat=24, mark phase=3,  mark options={solid}] table[x=papr_dB, y=beta_0,        col sep=comma] {data/fig3_PAPR_data.csv};
\addplot[color=cB13,   dotted,         thick, mark=o,        mark repeat=24, mark phase=7,  mark options={solid}] table[x=papr_dB, y=beta_1_3,      col sep=comma] {data/fig3_PAPR_data.csv};
\addplot[color=cB116,  densely dashed, thick, mark=asterisk, mark repeat=24, mark phase=15, mark options={solid}] table[x=papr_dB, y=beta_1_16,     col sep=comma] {data/fig3_PAPR_data.csv};
\addplot[color=cBpi,   densely dotted, thick, mark=x,        mark repeat=24, mark phase=21, mark options={solid}] table[x=papr_dB, y=beta_pi,       col sep=comma] {data/fig3_PAPR_data.csv};

\end{semilogyaxis}
\end{tikzpicture}